# Electronic phase separation in the rare earth manganates,

# $(La_{1-x}Ln_x)_{0.7}Ca_{0.3}MnO_3$ ($Ln$ = $Nd$, $Gd$ and $Y$)

**L. Sudheendra and C.N.R. Rao***

*Chemistry and Physics of Materials Unit, Jawaharlal Nehru Centre for Advanced Scientific Research, Jakkur P.O., Bangalore-560064, India.*

## Abstract

Electron transport and magnetic properties of three series of manganates of the formula $(La_{1-x}Ln_x)_{0.7}Ca_{0.3}MnO_3$ with $Ln = Nd$, $Gd$ and $Y$, wherein only the average A-site cation radius $< r_A >$ and associated disorder vary without affecting the $Mn^{4+}/Mn^{3+}$ ratio, have been investigated to understand the nature of phase separation. All the three series of manganates show saturation magnetization characteristic of ferromagnetism, with the ferromagnetic $T_c$ decreasing with increasing in $x$ up to a critical value of $x$, $x_c$ ($x_c$ = 0.6, 0.3, 0.2 respectively for Nd, Gd, Y). For $x > x_c$, the magnetic moments are considerably smaller showing a small increase around $T_M$, the value of $T_M$ decreasing slightly with increase in $x$ or decrease in $< r_A >$. The ferromagnetic compositions ($x \leq x_c$) show insulator-metal (IM) transitions, while the compositions with $x > x_c$ are insulating. The magnetic and electrical resistivity behavior of these manganates is consistent with the occurrence of phase separation in the compositions around $x_c$, corresponding to a critical average radius of the A-site cation, $< r_A^c >$, of 1.18 Å. Both $T_c$ and $T_{IM}$ increase linearly when $< r_A > > < r_A^c >$ or $x \leq x_c$ as expected of a homogenous ferromagnetic phase. Both $T_c$ and $T_M$ decrease linearly with the A-site cation size disorder at the A-site as measured by the variance $\sigma^2$. Thus, an increase in $\sigma^2$ favors the insulating AFM state. Percolative conduction is observed in the compositions with $< r_A > \; > \; < r_A^c >$. Electron transport properties in the insulating regime for $x > x_c$ conforms to the variable range hopping mechanism. More interestingly, when $x > x_c$, the real part of dielectric constant ($\varepsilon'$) reaches a high value ($10^4$-$10^6$) at ordinary temperatures dropping to a very small (~500) value below a certain temperature, the value of which decreases with decreasing frequency.

---

* *e-mail: cnrrao@jncasr.ac.in*

## Introduction

Rare earth manganates of the general formula $Ln_{1-x}A_xMnO_3$ ($Ln$ = rare earth, $A$ = alkaline earth) exhibit many interesting properties such as colossal magnetoresistance, charge ordering and electronic phase separation [1-3]. Phase separation in these materials has attracted much interest in the last few years because of the fascinating features associated with the phenomenon [3]. Of the various manganite compositions, the $(La_{1-x}Pr_x)_{0.7}Ca_{0.3}MnO_3$ system has provided really valuable information on the electronic phase separation [4-6]. This is a convenient system for the study of phase separation since the *Pr*-rich compositions are antiferromagnetic (*AFM*) and charge ordered (*CO*) where as the *La*-rich compositions are ferromagnetic metals (*FMM*). At a fixed *Ca* mole fraction of 0.3, a change in $x$ leads to a transition from *FMM* behavior to a charge ordered insulator (*COI*) behavior around a $x$ value of 0.7. That phase separation in $(La_{1-x}Pr_x)_{0.7}Ca_{0.3}MnO_3$ involves the *FMM* and *COI* domains, has been demonstrated by various types of measurements. This system exhibits several unusual features. Thus, in spite of the drop in resistivity at the insulator to metal (*IM*) transition, the *CO* peaks in the x-ray diffraction pattern continue to grow at low temperatures [7]. Besides the *COI* phase, another phase is suggested to be present below the charge ordering temperature [7]. Based on a neutron diffraction study, a phase diagram has been provided for this system wherein the $0.6 \leq x \leq 0.8$ region separates the homogeneous *FMM* and *AFM* phases [8].

Considering that the phase separation in $(La_{1-x}Pr_x)_{0.7}Ca_{0.3}MnO_3$ is essentially due to A-site cation disorder, we considered it important to investigate a series of compounds of the type $(La_{1-x}Ln_x)_{0.7}Ca_{0.3}MnO_3$ over a range of compositions $0 \leq x < 1$ where the variation in *Ln* brings about a marked change in the average A-site cation radius, $< r_A >$, as well as the size disorder. In this paper, we report the results of a systematic study of the electronic, magnetic and dielectric properties of three series of $(La_{1-x}Ln_x)_{0.7}Ca_{0.3}MnO_3$ with *Ln = Nd, Gd* and *Y*. It is to be noted that in all the compositions of the three series of manganates, the $Mn^{4+}/Mn^{3+}$ ratio remains constant, the only variable being the average A-site cation radius and the associated disorder.

**Experimental**

Compositions of the formula $(La_{1-x}Ln_x)_{0.7}Ca_{0.3}MnO_3$ (*Ln = Nd, Gd* or *Y*) were prepared by standard solid-state synthesis. A stoichiometric mixture of the rare earth acetate, $CaCO_3$ and $MnO_2$ were mixed thoroughly in an agate mortar with the help of iso-propyl alcohol, and the mixture was first fired to 1273 K for 48 h with two intermediate grindings. This preheated sample was then further heated at 1473 K for 24 h, pressed into pellets and fired to 1648 K for 12 h to obtain the single-phase compounds. The samples were tested for phase purity by X-ray analysis using Seifert 3000 diffractometer and the composition was checked by EDAX measurements. All the compositions were fitted to an orthorhombic unit cell with *Pnma* space group. The unit cell dimensions of $(La_{1-x}Ln_x)_{0.7}Ca_{0.3}MnO_3$ decreased with increase in *x* in all the three series of compounds (Fig.1), with a corresponding decrease in the unit cell volume as expected when a large cation is replaced by smaller cation. The dependence of the unit cell dimensions and volume for three different compositions parallels the dependence of the average radius of the A-site cations, $< r_A >$, on $x$. Thus a plot of the unit cell volume of all the series against $< r_A >$ is linear as shown in the inset of Fig. 1,

Magnetization measurements were carried out in a Lakeshore vibrating sample magnetometer in the 300-50 K temperature range at an applied field of 4000 G. Electrical transport properties were measured by standard four-probe method with silver epoxy as the electrodes in the 300-20 K temperature range. Dielectric measurements were carried out with the help of Agilent 4294A impedance analyzer with gold coating on either side of the disk shaped samples acting as electrodes. The data was collected from 85-300 K.

**Results and discussion**

In Fig.2a, we show the results of magnetic measurements of the $(La_{1-x}Nd_x)_{0.7}Ca_{0.3}MnO_3$ series to show how the magnetic moment ($\mu_\beta$) is sensitive to the substitution of the smaller $Nd^{3+}$ cation in place of $La^{3+}$. We see a clear *FM* transition down to $x = 0.5$ with a saturation magnetic moment close to $3\mu_\beta$. The ferromagnetic Curie temperature, $T_c$, shifts to lower temperatures with increase in $x$. We fail to see magnetic saturation in compositions with $x \geq 0.6$ and instead, the maximum value of $\mu_\beta$ is far less than three at low temperatures. We designate the composition up to which ferromagnetism occurs as the critical composition $x_c$. The compositions with $x > x_c$ show a gradual, definitive increase in the magnetization or $\mu_\beta$ at a temperature $T_M$, the $T_M$ value decreasing with increasing $x$. It is possible that $T_M$ represents the onset of canted antiferromagnetism (*CAF*). In $(La_{1-x}Gd_x)_{0.7}Ca_{0.3}MnO_3$, ferromagnetism is observed up to $x = 0.3$. The $T_c$ decreases with increase in $x$ in the composition range $0.0 \leq x \leq 0.3$. For $x > 0.3$, we observe a gradual increase in $\mu_\beta$ value at low temperature around $T_M$, with the value of $T_M$ decreasing with increase in $x$. Note that the value of $x_c$ (~0.3) in the $Gd$ series is considerably lower than that in the $Nd$ series where it was 0.6, showing that $x_c$ decreases with the decrease in the average radius of the A-site cation, $< r_A >$. The results of the $(La_{1-x}Gd_x)_{0.7}Ca_{0.3}MnO_3$ series of manganates (Fig.2b) obtained by us agree with those reported for the $x$ = 0.0-0.25 compositions by Terashita and Neumeier [9]. In $(La_{1-x}Y_x)_{0.7}Ca_{0.3}MnO_3$, ferromagnetism is seen only for compositions with $x \leq 0.2$, the $T_c$ decreasing with increase in $x$. Compositions with $x > 0.2$, show a gradual increase in the $\mu_\beta$ value below $T_M$ and the $T_M$ decreases with increase in $x$ (Fig. 2c). Thus, the $x_c$ value in $(La_{1-x}Ln_x)_{0.7}Ca_{0.3}MnO_3$ is 0.75, 0.6, 0.3 and 0.2 for $Ln$ = $Pr$, $Nd$, $Gd$ and $Y$ respectively, showing a sensitive dependence of $x_c$ on $< r_A >$.

The magnetization data of $(La_{1-x}Ln_x)_{0.7}Ca_{0.3}MnO_3$ with $Ln$ = $Nd$, $Gd$ and $Y$ in the composition range $x \geq x_c$ show certain features of significance . Thus, we find that in the compositions close to $x_c$ ($x$ ~ 0.6-0.7 when $Ln$ = $Nd$; $x$ ~ 0.3-0.4 in the case of $Gd$), the low temperature $\mu_\beta$ value is significant, reaching values anyway between 1-2.5. It is only when $x$ is considerably large ($x >> x_c$), the $\mu_\beta$ value decreases to values less than unity. These relatively high values of $\mu_\beta$, when $Ln$ = $Nd$ and $Gd$, are probably due to moments of the rare earths. When $Ln$ = $Y$, the $\mu_\beta$ values are all low, the highest value of ~ $1\mu_\beta$ being observed when $x$ ~ $x_c$ (Fig. 2c). When $x > x_c$, the magnetic moment decreases below $0.5\mu_\beta$. When $x >> x_c$, the presence of small $FM$ clusters cannot be ruled out, even though the $CAF$ interactions may be dominant. The drastic change in the values of $\mu_\beta$ around $x_c$ in the three series of manganates can be attributed to phase separation due to disorder caused by substitution of the smaller rare earth cations in place of *La*. In the *Pr* system, phase separation have been reported in the regime of $x \sim x_c$ ($x \sim 0.6$-$0.8$) [8].

In Fig. 3, we have plotted the ferromagnetic $T_c$ of the $(La_{1-x}Ln_x)_{0.7}Ca_{0.3}MnO_3$ composition against $x$ ($x \leq x_c$). The $T_c$ decreases with increase in $x$ linearly. In Fig. 3, we have also plotted the $T_M$ values of the non-ferromagnetic compositions ($x > x_c$), to show how the $T_M$ value also decreases with increase in $x$, albeit with a considerably smaller slope. In the inset of Fig. 3, we have plotted the $T_c$ and $T_M$ values against $< r_A >$. The $T_c$ value increases with increase in $< r_A >$ for $< r_A > \geq 1.18$ Å (or $x < x_c$), while $T_M$ increases with $< r_A >$ with a smaller slope. The $< r_A >$ value of 1.18 Å marks the critical radius beyond which ferromagnetism manifests itself in $(La_{1-x}Ln_x)_{0.7}Ca_{0.3}MnO_3$. It must be noted that in a variety of manganates of the general compositions $Ln_{1-x}A_xMnO_3$, an $< r_A >$ of 1.18 Å marks the critical value below which charge ordering becomes robust, rendering it difficult to destroy it by applying magnetic fields or impurity substitutions [10]. An $< r_A >$ 1.18 Å also marks the critical value below which reentrant *FM* transitions with $T_c < T_{co}$ are seen [11]. The compositions with $< r_A > < 1.18$ Å are *AFM* insulators. In Fig. 4, we have plotted $\mu_\beta$ against $< r_A >$ for the three series of $(La_{1-x}Ln_x)_{0.7}Ca_{0.3}MnO_3$. The $\mu_\beta$ values when $< r_A > < 1.18$ Å are rather small, well below $3\mu_\beta$, generally less than $1\mu_\beta$. Magnetization in this regime arises from *FM* interactions in a primarily *AFM* environment probably due to the presence of magnetic clusters. Thus, the $\mu_\beta$ values in the range of 1-2.5$\mu_\beta$ in Fig. 4 correspond to the phase separation regime where the $< r_A >$ values are in the 1.18-1.165 Å range. This is consistent with results reported for $(La_{1-x}Pr_x)_{0.7}Ca_{0.3}MnO_3$ system [8]. When $< r_A > \leq 1.165$ Å, the $\mu_\beta$ values become considerably smaller, although ferromagnetic interactions due to clusters are likely to be present at low temperatures in this regime as well.

The electrical resistivity data of $(La_{1-x}Ln_x)_{0.7}Ca_{0.3}MnO_3$ ($Ln$ = $Nd$, $Gd$ and $Y$) reflect the magnetization data, with the $x \leq x_c$ compositions showing *IM* transitions (Fig. 5). The compositions with $x > x_c$ are insulating and do not exhibit *IM* transitions. For $x \leq x_c$, the value of the resistivity at the *IM* transition increases with the increase in $x$, and a change in the resistivity of 3-4 orders magnitude is observed at the transition. Thus, the value of resistivity for at 20 K for $x \approx x_c$ is considerably higher than that of $La_{0.7}Ca_{0.3}MnO_3$.The temperature of the *IM* transition, $T_{IM}$, in $x \leq x_c$ compositions decreases linearly with increasing $x$ (Fig.6). The $T_{IM}$ versus $< r_A >$ plot is linear with a positive slope as expected (see inset of Fig. 6)

The small but finite magnetic moments and relatively large resistivities at low temperatures found in $(Ln_{1-x}Ln_x)_{0.7}Ca_{0.3}MnO_3$ for $Ln$ = $Pr$, $Nd$, $Gd$ and $Y$ around $x_c$ or $< r_A^c >$ are a consequence of phase separation. Phase separation also causes thermal hysteresis in the resistivity behavior around the *IM* transitions (see the insets in Fig. 5). The insets in figures 5a, b and c show that the resistivity in the warming cycle is lower than that in the cooling cycle up to a certain temperature beyond which the resistivities in the two cycles merge. Upon cooling the sample below the *IM* transition, the *FMM* phase grows at the expense of *AFM* insulating phase causing a decrease in the resistivity value. When the same sample is warmed, the insulating phase grows at the expense of the *FMM* phase, the latter providing the conductive path. The thermal hysteresis in resistivity is therefore due to percolative conductivity in these manganates, the hysteresis decreasing with the increase in $< r_A >$ or decrease in $x$ as expected. It must be noted that the compositions with $< r_A > >$1.18 Å which are in a homogenous *FM* phase do not show any hysteresis around $T_{IM}$. Furthermore, the ratio of the peak resistivities in the cooling and heating cycles ($\rho_c/\rho_w$) is constant for samples with $< r_A > >$ 1.18 Å in the homogenous *FM* regime, but increases significantly with decrease in $< r_A >$.

Since phase separation in the $(La_{1-x}Ln_x)_{0.7}Ca_{0.3}MnO_3$ series of manganates mainly arises from size disorder caused by the substitution of small ions in place of *La*, we wanted to quantitatively examine the effect of size disorder in these manganates, in terms of the size variance of the A-cation radius distribution, $\sigma^2$ [12,13]. For two or more A-site species with fractional occupancies $x_i$ ($\sum x_i = 1$), the variance of the ionic radii $r_i$ about the mean $< r_A >$ is given by $\sigma^2 = (\sum x_i r_i^2 - <r_A>^2)$. We have examined the electrical and magnetic properties for a series of $(La_{1-x}Ln_x)_{0.7}Ca_{0.3}MnO_3$ with fixed $< r_A >$ and variable $\sigma^2$. In fig.7, we have shown typical plots of temperature variation of $\mu_\beta$ for fixed $< r_A >$ values of 1.196 and 1.17 Å. We see a decrease in $T_c$ with increase in $\sigma^2$. For $< r_A > =$ 1.17 Å the material is not ferromagnetic. The $\mu_\beta$ value as well as $T_M$ show a marked decrease with increase in $\sigma^2$. The resistivity decreases with an increase in magnetization, and when $< r_A > =$ 1.196 Å the $T_{IM}$ decreases with increase in $\sigma^2$. In Fig. 8, we have plotted $T_c$ values of the manganates against $\sigma^2$ for compositions with fixed $< r_A >$ values > 1.18 Å along with the $T_M$ value for fixed $< r_A >$ values ≤ 1.18 Å. Surprisingly, the slopes of the $T_c$ and $T_M$ plots are similar (~9000 K/Å$^2$). The slope of $T_c$ versus $\sigma^2$ in $Ln_{0.5-x}La_xCa_{0.5}MnO_3$, where *Ln* = *Nd* and *Pr*, compositions is reported to be ~ 15000 K/Å$^2$ [11]. The similarity in behavior of these different compositions may be due to presence of ferromagnetic correlations even when $< r_A > \leq < r_A^c >$ or $x > x_c$. By extrapolating the $T_c$ and $T_M$ values to $\sigma^2 = 0$, we obtain the intercepts $T_c^o$ and $T_M^o$, which represent the transition temperatures in the absence of any size disorder. We have plotted the $T_c^o$ and $T_M^o$ values against $< r_A >$ in the inset of Fig. 8. The $T_c^o$ and $T_M^o$ values increase with increase in $< r_A >$ as expected.

As can be seen from Fig. 5, the resistivity of $(La_{1-x}Ln_x)_{0.7}Ca_{0.3}MnO_3$ compositions show an a IM transition up to $x_c$, and these compositions show a significant increase in resistivity with increase in $x$ more prominently than the compositions with $x > x_c$. Electrical conductivity in the $x \approx x_c$ compositions, are percolative at low temperatures and accordingly, we are able to fit the low temperature data for the compositions with $< r_A >$ > 1.18 Å to a percolative scaling law, $log\rho \propto \log| < r_A > - < r_A^c > |$ as shown in Fig.9. The slope of the plot is -2.63, a value close to the experimental and predicted values for percolative systems [3].

We have examined the resistivity data of the insulating compositions with $x > x_c$ in some detail to explore whether they conform to activated hopping defined by $log\rho \propto 1/T^n$ where, $n$ = 1, 2 or 4. Here, $n$ = 1 corresponds to a simple Arrhenius conductivity. When $n$ = 2, the hopping is called nearest neighbor hopping (NNH). The hopping is controlled by Coulombic forces. When $n$ = 4, the hopping is termed variable range hopping (VRH) and the hopping dynamics is controlled by collective excitation of the charge carries. Figure 10 shows typical fits of the resistivity data of $(La_{1-}$ ${}_{x}Y_{x})_{0.7}Ca_{0.3}MnO_3$ for $x > x_c$. The resistivity data could be fitted to $T^{-1/2}$ dependence with standard deviations varying from 0.008-0.04. The standard deviations for the $T^{-1/4}$ fits were 0.02-0.07. AC conductivity ($\sigma_{ac}$) measurements show a frequency dependence of conductivity, the conductivity increasing with increasing frequency, ω, and decreasing $x$ for a given $Ln$ cation. A fit of $\sigma_{ac}$ to $\omega^s$, gives a value of $s$ in the range 0.5-0.8, consistent with the VRH mechanism [14]. At high temperatures (>150 K), the conductivity exhibits a frequency independence indicating the dominance of DC conductivity.

In Fig. 11, we show the temperature response of the real part of the dielectric constant ($\varepsilon'$) for $(La_{0.3}Y_{0.7})_{0.7}Ca_{0.3}MnO_3$ at different frequencies. The $\varepsilon'$ reaches a high values at ordinary temperatures but decreases dramatically below ~120 K. The temperature at which the drop in $\varepsilon'$ occurs increases with increasing frequency, exhibiting a relaxor type behavior. The dielectric relaxation obtained in the $(La_{1-x}Ln_x)_{0.7}Ca_{0.3}MnO_3$ compositions is similar to that found in copper titanate perovskite dielectrics [15]. The unusually high dielectric constant of $(La_{1-x}Ln_x)_{0.7}Ca_{0.3}MnO_3$ for $x > x_c$ can be understood by assuming the presence of conducting domains surrounded by insulating layers with an activated behavior of the intra-domain (non-percolative) conductivity. Such a finite conductivity could be one of the reasons for the high dielectric constant beyond 120 K. The dielectric constant at any given frequency decreases with the decrease in conductivity for $x > x_c$ as expected [16]. Similar to the copper titanates, the $(La_{1-x}Ln_x)_{0.7}Ca_{0.3}MnO_3$ ($x > x_c$) compositions exhibit peaks in the dissipation factor which shifts to higher frequencies with increasing temperature. An Arrhenius plot of frequency with temperature gives a activation energy of 68 meV close to the values in perovskite oxide dielectrics.

## Conclusions

The present study of the electronic and magnetic properties of three series of rare earth manganates of the type $(La_{1-x}Ln_x)_{0.7}Ca_{0.3}MnO_3$ where $Ln = Nd, Gd$ and $Y$, has shown they become ferromagnetic when $x \leq x_c$ accompanied by an insulator-metal transition. These manganates become AFM insulators when $x > x_c$, but show a small increase in magnetic moment at low temperatures ($T < T_M$). $T_c$ and $T_M$ values are both sensitive to cation size disorder. The values of $x_c$ are 0.6, 0.3 and 0.2 respectively for *Nd*, *Gd* and *Y*, correspond to a unique value of average size of the A-site cation, $< r_A >$, of 1.18 Å ($< r_A^c >$). It must be recalled that this $< r_A >$ value also marks the critical value below which the ground state is charge ordered. Phase separation is marked around $x_c$ or $< r_A^c >$ in all the series of compositions, although ferromagnetic clusters are likely to be present at low temperatures even when $1.0 > x > x_c$. The origin of phase separation in these series of manganates is attributed to the size disorder arising from the substitution of the smaller rare earth cations in place of La.

## Acknowledgment

The authors thank BRNS (DAE) for support of this research.

**References**

1) Colossal Magnetoresistance, Charge Ordering and Related Properties of Manganese Oxides, edited by C.N.R. Rao ad B. Raveau (World Scientific, Singapore, 1998).

2) Colossal Magnetoresistance Oxides, edited by Y. Tokura (Gordon & Breach, London, 1999).

3). C.N.R. Rao and P.V. Vanitha, Curr. Opp. Solid State Mater. Sci., 6, 97, 2002.

4) M. Uehara, S. Mori, C.H. Chen and S.-W. Cheong, Nature, 399, 560, 1999.

5) N.A. Babushkina, A.N. Taldenkov, L.M. Belova, E.A. Chistotina, O. Yu. Gorbenko, A.R. Kaul, K.I. Kugel and D.I. Khomskii, Phys. Rev., B62, R6081, 2000.

6) H.J. Lee, K.H. Kim, M.W. Kim, T.W. Noh, B.G.Kim, T.Y. Koo, S.-W. Cheong, Y.J. Wang and X. Wei, Phys Rev., B65, 115118,2002.

7) V. Kiryukhin, B.G. Kim, V.Podzorov, S.-W. Cheong, T.Y. Koo, J.P. Hill, I. Moon and Y. H. Jeong, Phys. Rev., B63, 024420, 2000.

8) A.M. Balagurov, Y.Yu. Pomjakushin, D.V. Sheptyakov, V.L. Aksenov, P. Fischer, L. Keller, O. Yu. Gorbenko, A.R. Kaul and N.A. Babushkina, Phys Rev., B64, 024420, 2001.

9) H. Terashita and J.J. Neumeier, Phys. Rev., B63, 174436, 2001.

10) C.N.R. Rao, A. Arulraj, A.K. Cheetam andB. Raveau, J.Phys.: Condens. Matter, R83, 12, R 83, 2000.

11) P.V. Vanitha and C.N.R. Rao, J. Phys.: Condens Matter., 13, 11707, 2001.

12) L. M. Rodriguez-Martinez and J.P. Attfield, Phys Rev., B54, R15622, 1996.

13) P.V. Vanitha, P.N. Santosh, R.S. Singh, C.N.R. Rao and J.P. Attfield, Phys Rev., B59, 13539, 1999.

14) K.V. Vijaya Sarathy, S. Parashar, A.R. Raju and C.N.R. Rao, Solid State Sci., 4, 353, 2002.

15) C.C. Homes, T. Vogt, S.M. Shaprio, S. Wakimoto and A. R. Ramirez, Science, 293, 673, 2001.

16) L. He, J.B. Neaton, M.H. Cohen, D. Vanderbilt and C.C. Homes, Cond-mat/0110166, 2002.

**Figure captions**

Fig.1 Variation of unit cell dimension $a$ (Å) with $x$ in $(La_{1-x}Ln_x)_{0.7}Ca_{0.3}MnO_3$. Inset shows the variation of unit cell volume with $< r_A >$ (Å) in $(La_{1-x}Ln_x)_{0.7}Ca_{0.3}MnO_3$.

Fig.2 Temperature variation of $\mu_\beta$ in (a) $(La_{1-x}Nd_x)_{0.7}Ca_{0.3}MnO_3$, (b) $(La_{1-x}Gd_x)_{0.7}Ca_{0.3}MnO_3$ and (c) $(La_{1-x}Y_x)_{0.7}Ca_{0.3}MnO_3$.

Fig.3 Variation of $T_c$ with $x$ of $(La_{1-x}Ln_x)_{0.7}Ca_{0.3}MnO_3$. Solid symbols represent real $T_c$ and broken symbols represent $T_M$. Inset shows the variation of $T_c$ with $< r_A >$ (Å) for $(La_{1-x}Ln_x)_{0.7}Ca_{0.3}MnO_3$ ($x=0$ data is shown as ♦). $(La_{1-x}Ln_x)_{0.7}Ca_{0.3}MnO_3$ data taken from ref. 8.

Fig.4 Variation of $\mu_\beta$ with $< r_A >$ (Å) in $(La_{1-x}Ln_x)_{0.7}Ca_{0.3}MnO_3$ at 50 K.

Fig.5 Temperature variation of resistivity in (a) $(La_{1-x}Nd_x)_{0.7}Ca_{0.3}MnO_3$, (b) $(La_{1-x}Gd_x)_{0.7}Ca_{0.3}MnO_3$ and (c) $(La_{1-x}Y_x)_{0.7}Ca_{0.3}MnO_3$. The broken lines in the inset show warming cycle data.

Fig.6 Variation of $T_{IM}$ with $x$ in $(La_{1-x}Ln_x)_{0.7}Ca_{0.3}MnO_3$ for $x \le x_c$. Inset shows the variation of $T_{IM}$ with $< r_A >$ (Å) for the same composition range.

Fig.7 Temperature variation of $\mu_\beta$ in $(La_{1-x}Ln_x)_{0.7}Ca_{0.3}MnO_3$ a) $< r_A > =1.196$ Å ; b) $< r_A > =1.17$ Å. The inset shows the corresponding variations in resistivities.

Fig.8 Variation of $T_c$ (solid symbol) and $T_M$ (broken symbol) against $\sigma^2$ (Å$^2$). Inset shows the variation of $T_c^o$ and $T_M^o$ against $< r_A >$ (Å).

Fig.9 Linear scaling of $log\rho$ with $log|< r_A > - < r_A^c >|$ in $(La_{1-x}Ln_x)_{0.7}Ca_{0.3}MnO_3$ at 20 and 50 K.

Fig.10 Fits of the resisitivity data of $(La_{1-x}Y_x)_{0.7}Ca_{0.3}MnO_3$ for $x > x_c$ to (a) $T^{-1/2}$ and (b) $T^{-1/4}$ laws in the 90-220 K range. The open symbols are experimental data points and broken lines represents the corresponding linear fits.

Fig.11 Temperature variation of real part of dielectric constant ($\varepsilon'$) of $(La_{0.3}Y_{0.7})_{0.7}Ca_{0.3}MnO_3$ at different frequencies.

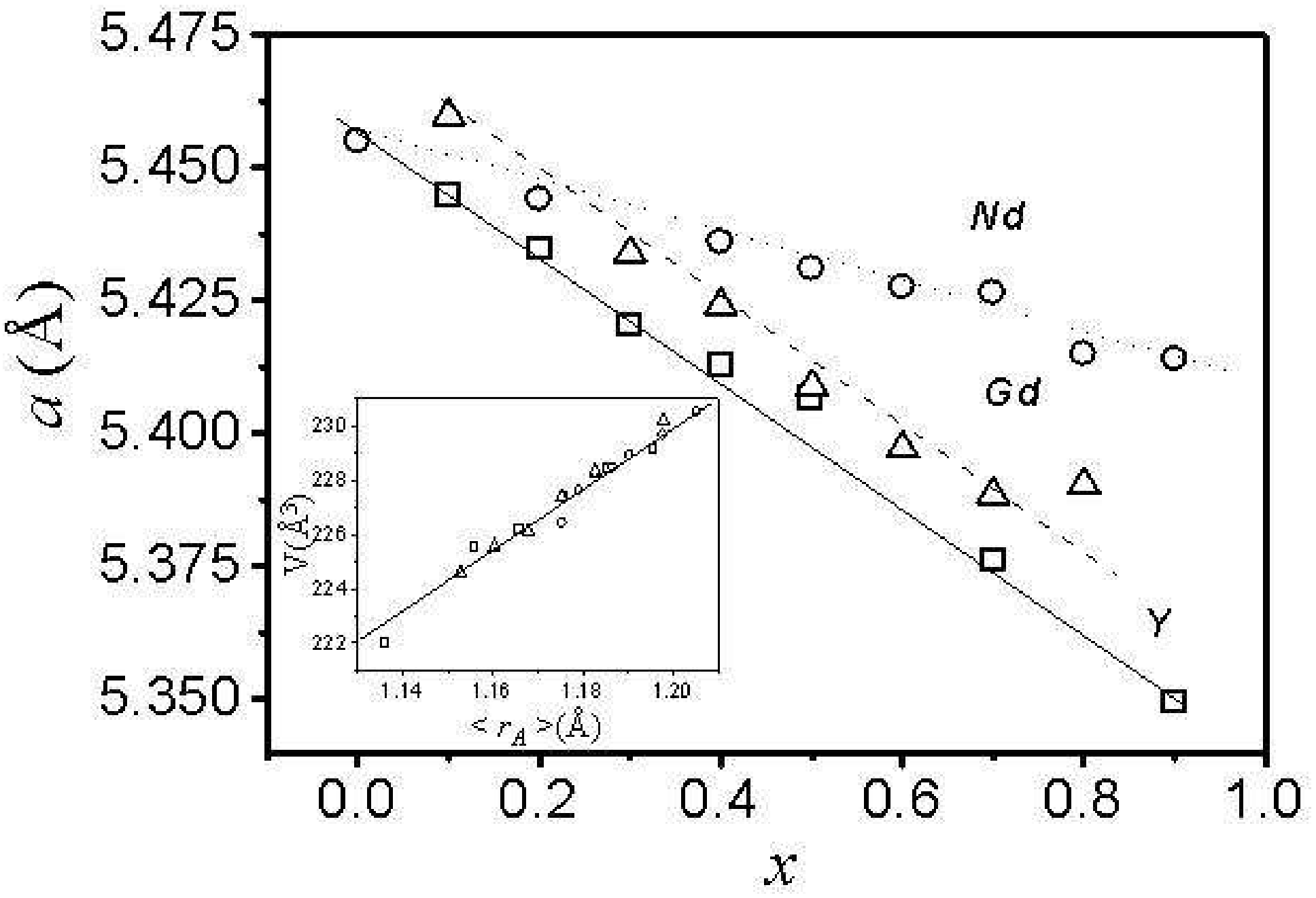

**Fig. 1** *L. Sudheendra and C.N.R. Rao*

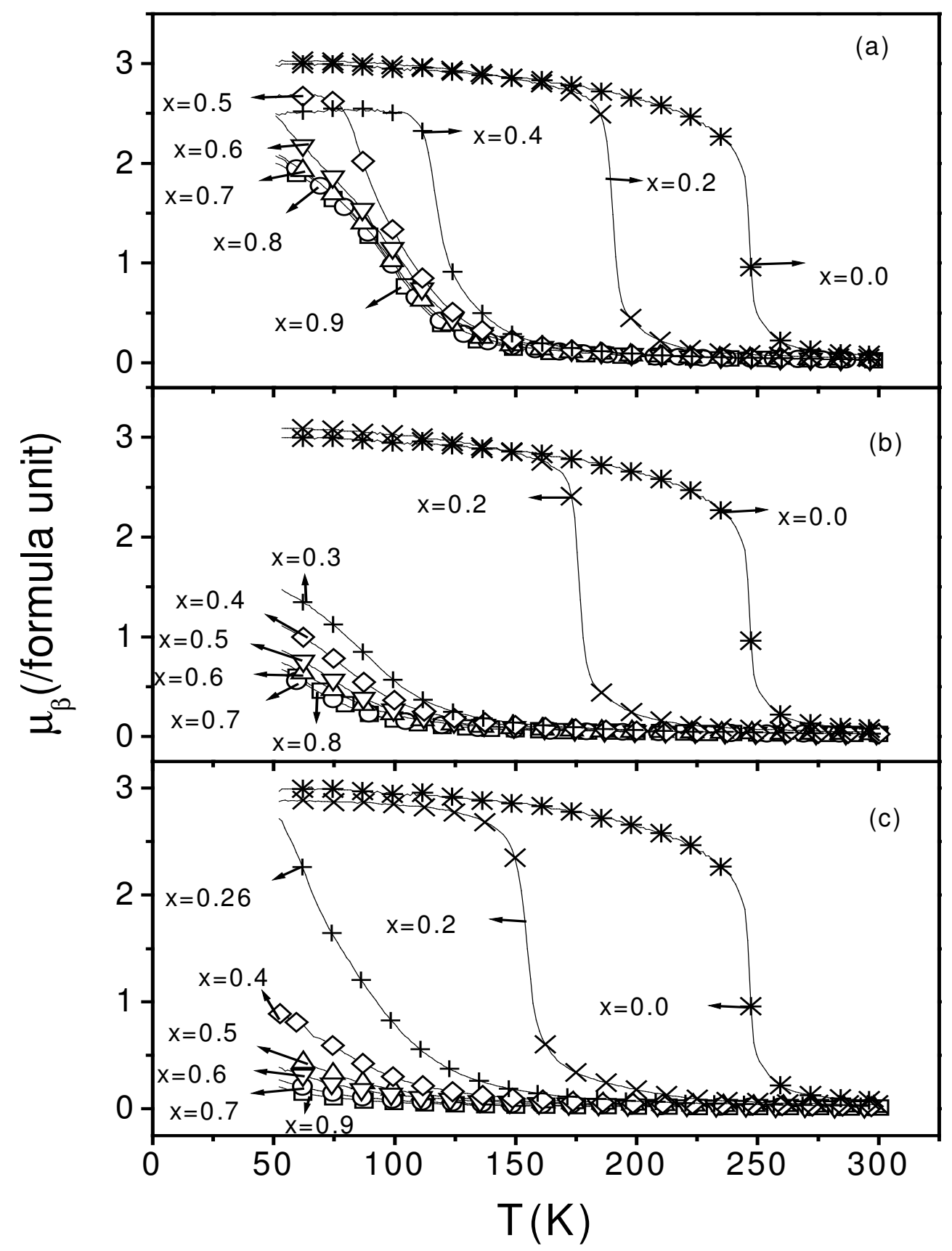

**Fig. 2** ***L. Sudheendra and C.N.R. Rao***

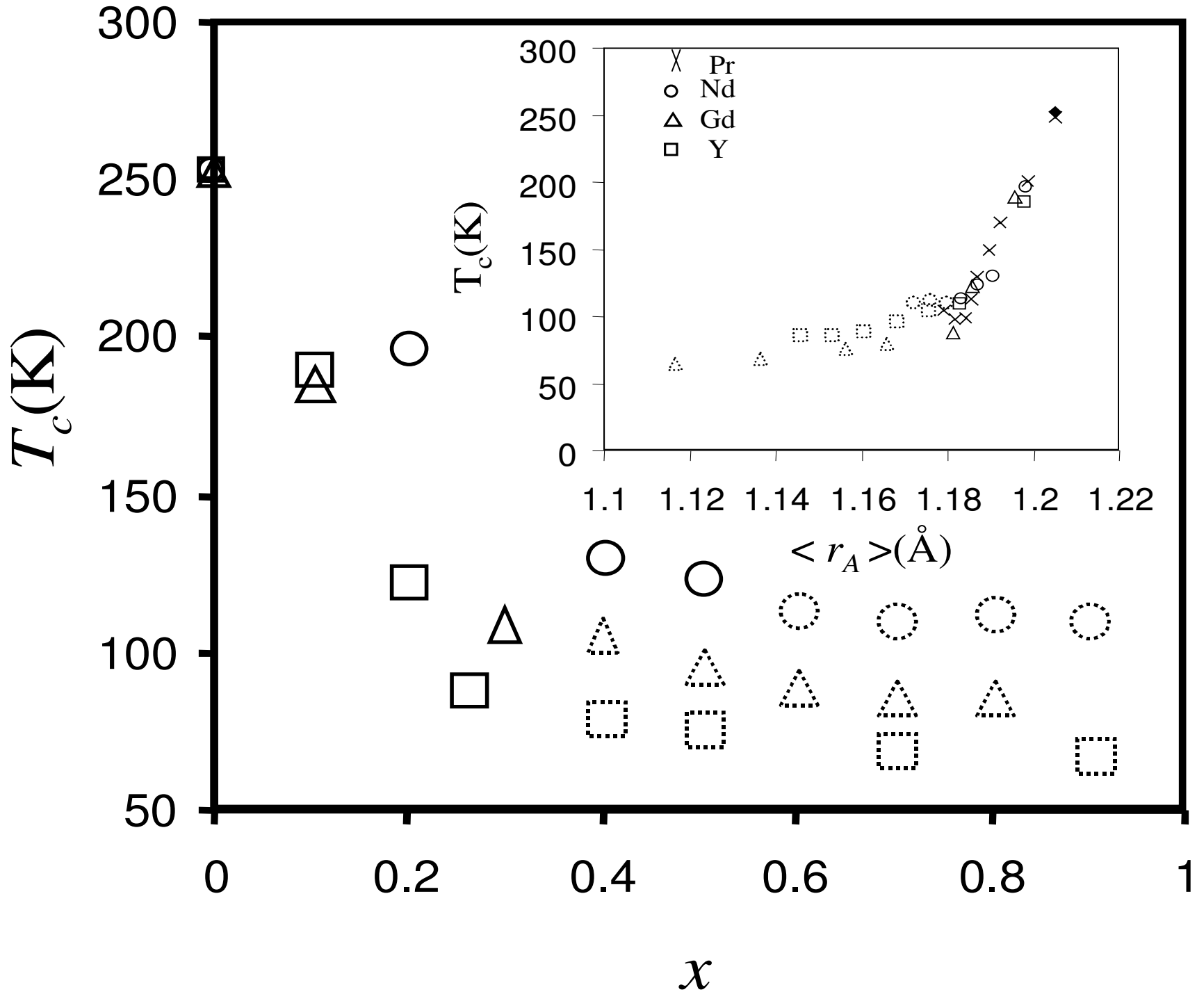

**Fig. 3** ***L. Sudheendra and C.N.R. Rao***

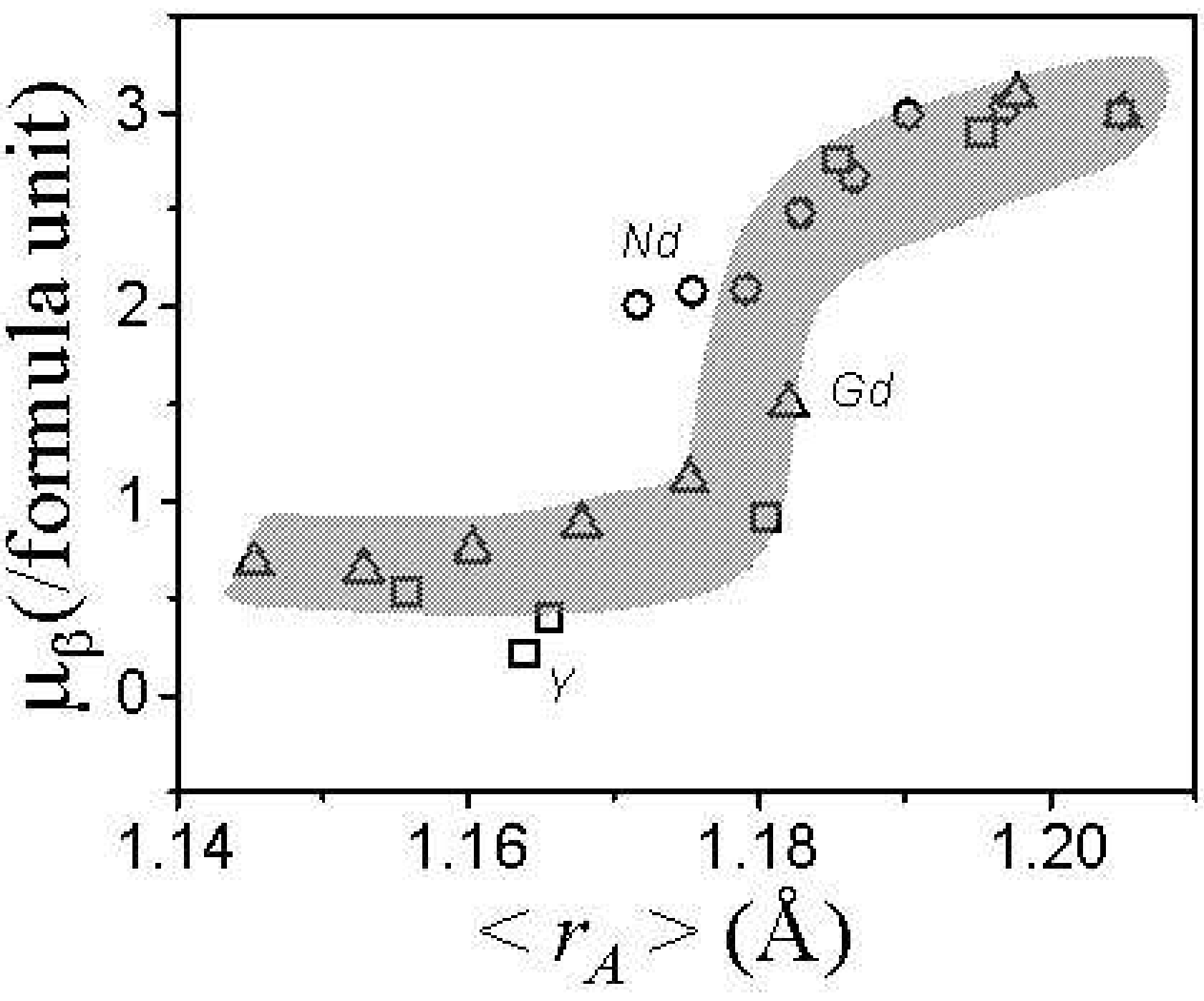

**Fig. 4** ***L. Sudheendra and C.N.R. Rao***

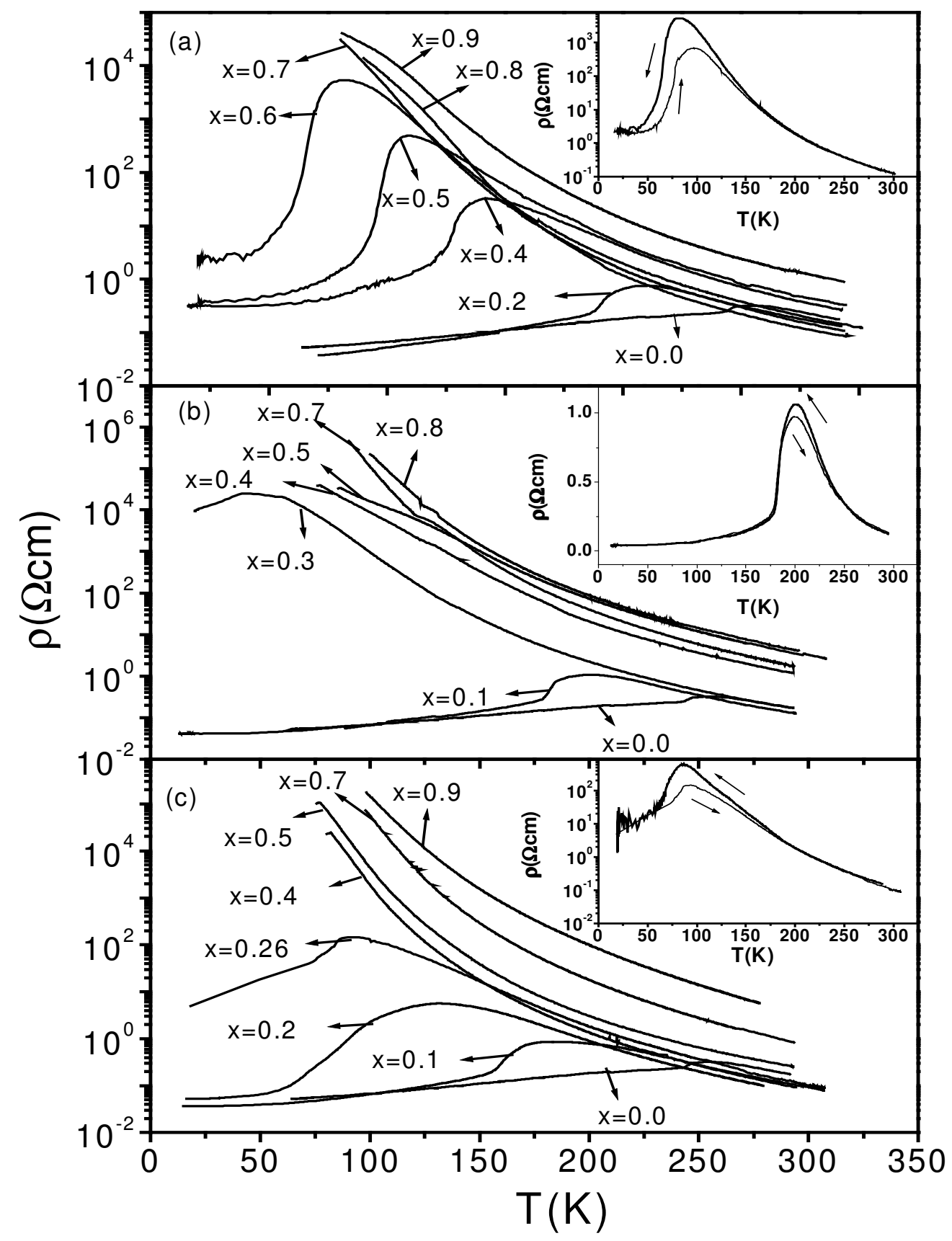

**Fig. 5** ***L. Sudheendra and C.N.R. Rao***

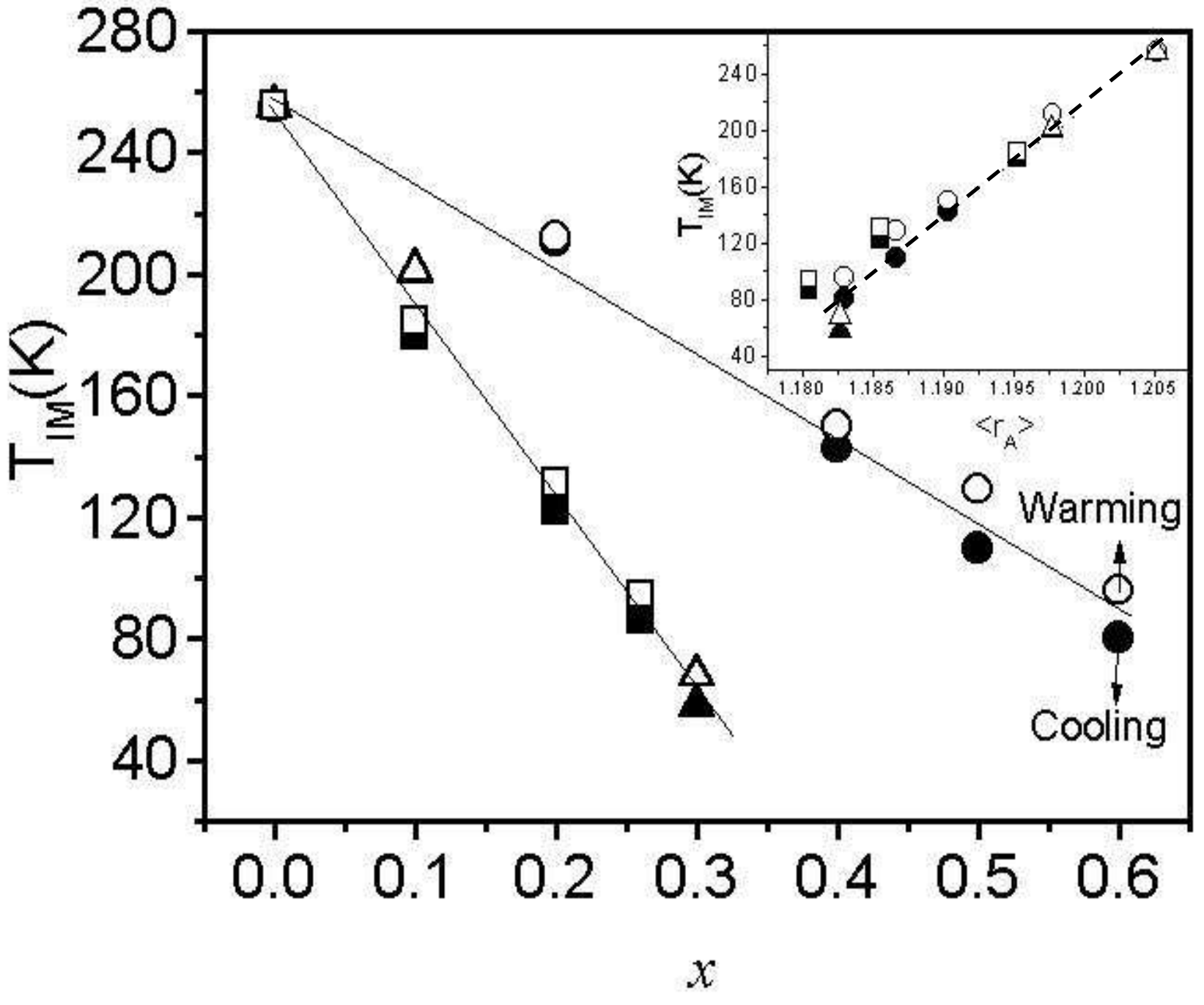

**Fig. 6** ***L. Sudheendra and C.N.R. Rao***

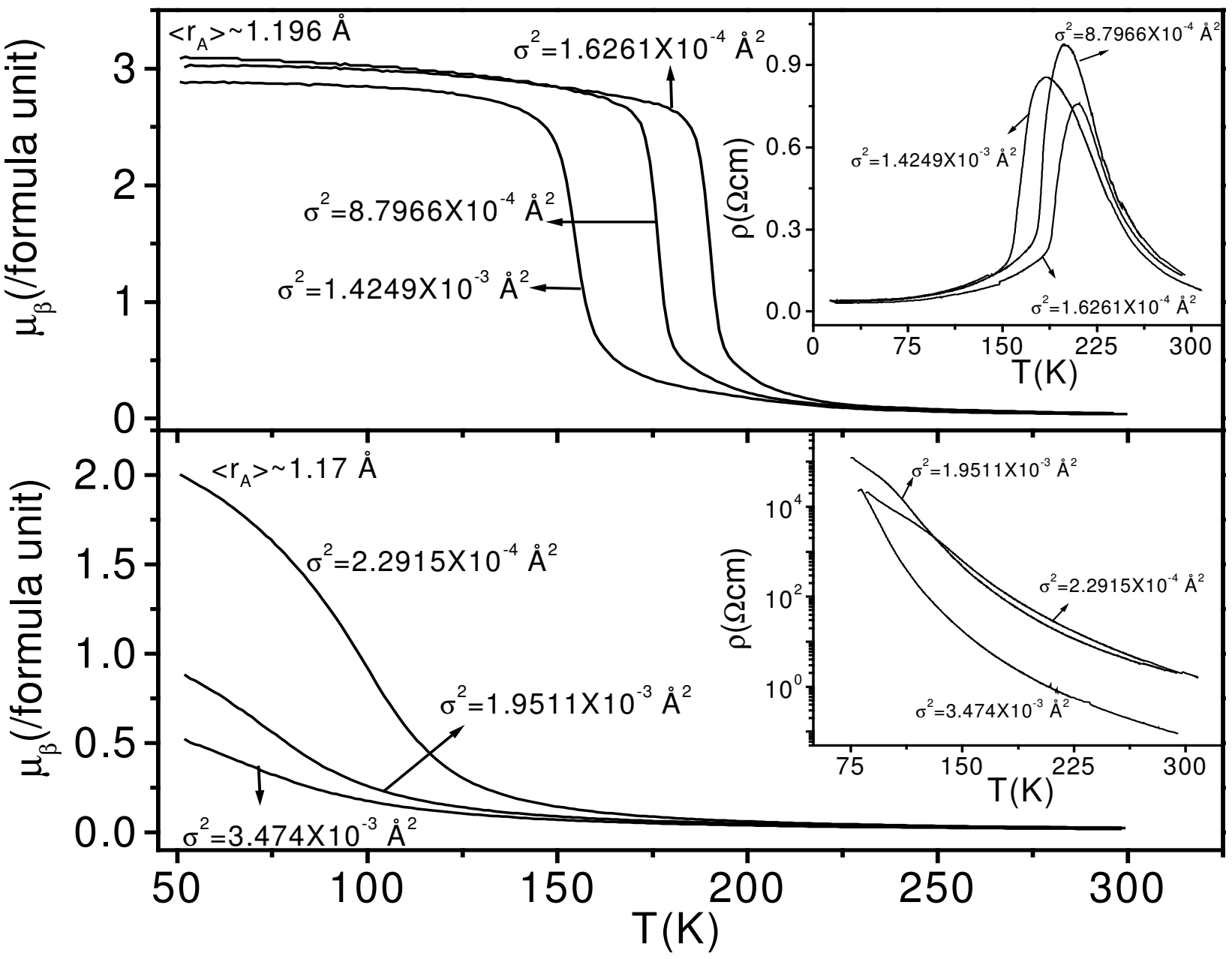

**Fig. 7** ***L. Sudheendra and C.N.R. Rao***

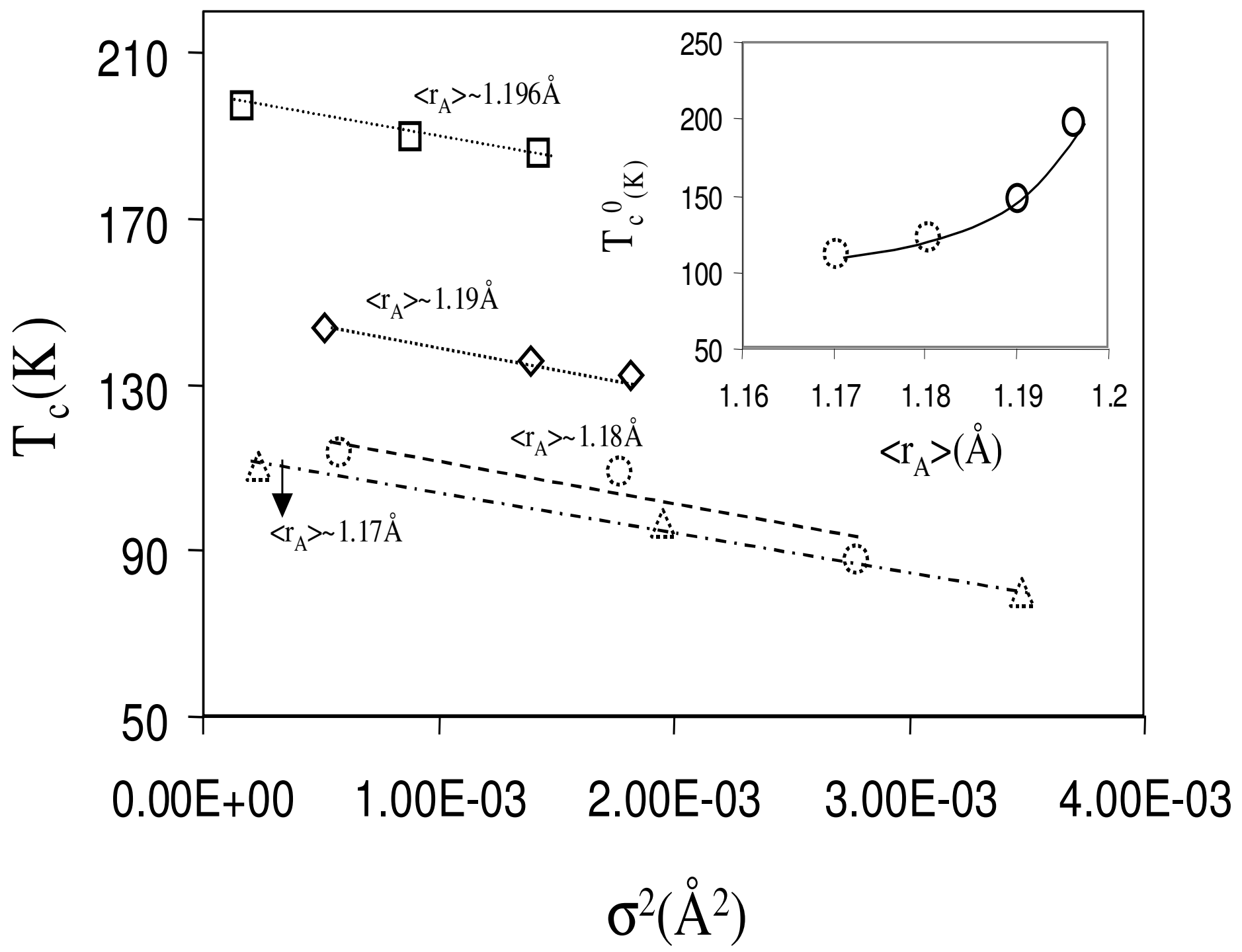

**Fig. 8** ***L. Sudheendra and C.N.R. Rao***

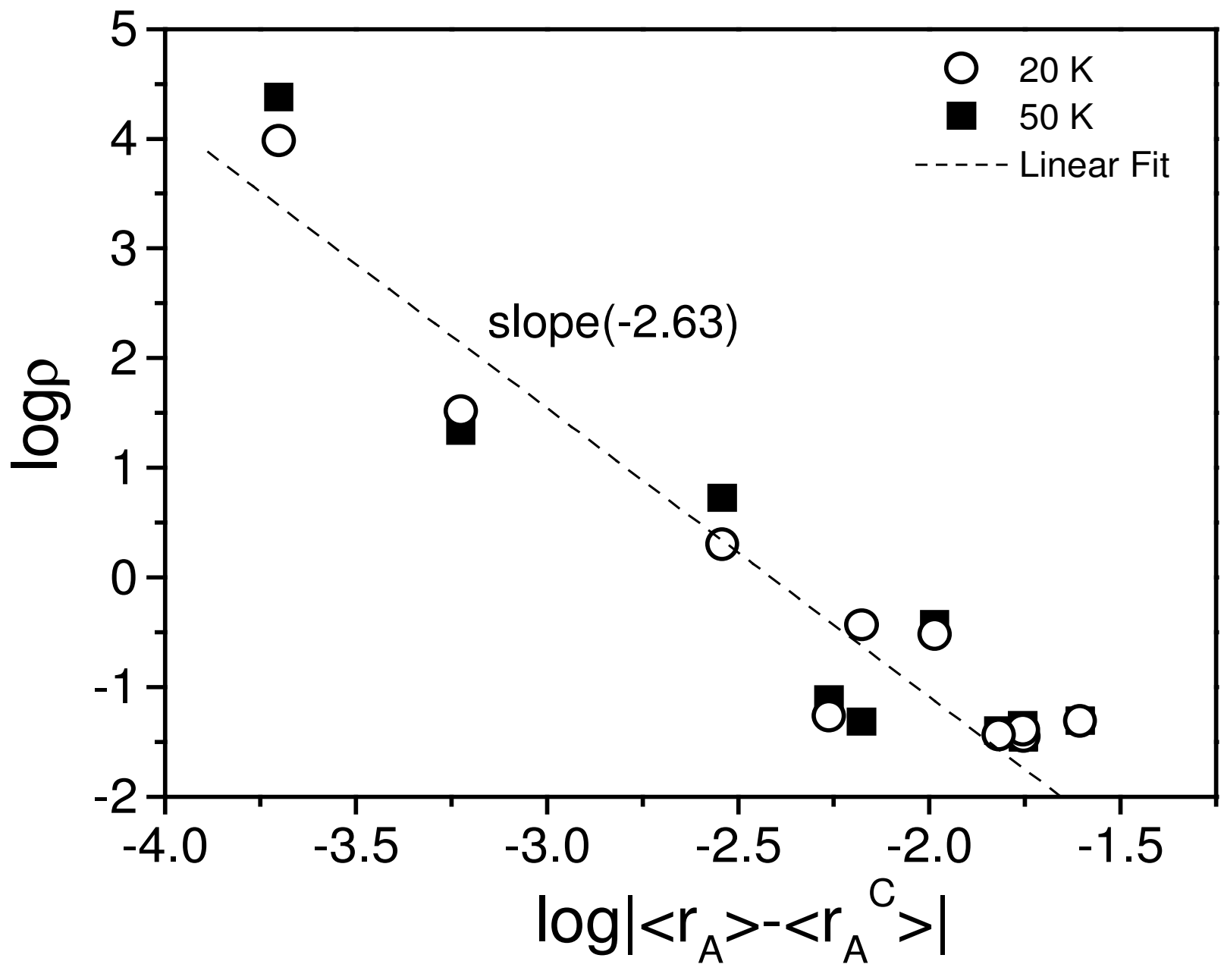

**Fig. 9** ***L. Sudheendra and C.N.R. Rao***

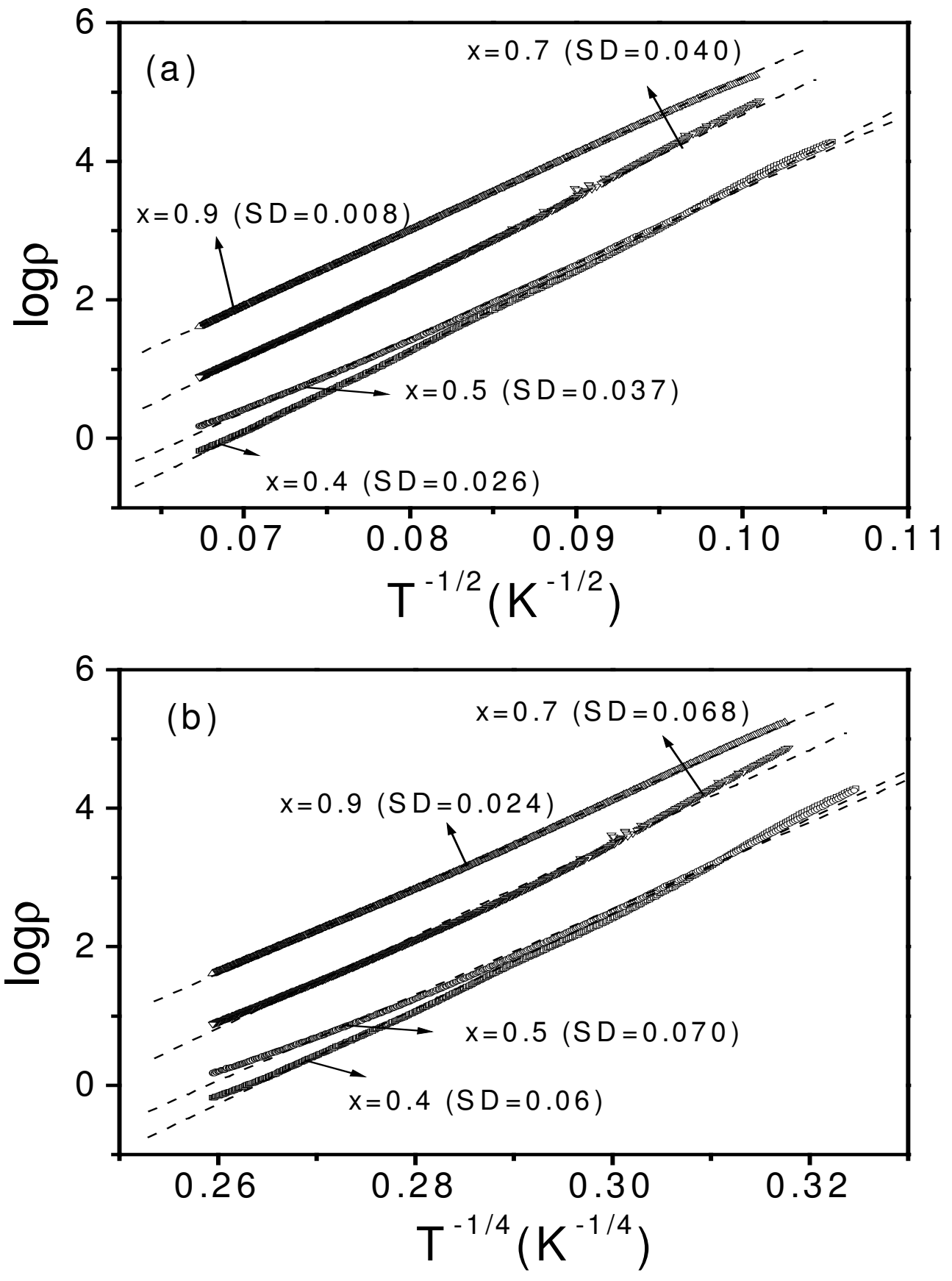

**Fig. 10** ***L. Sudheendra and C.N.R. Rao***

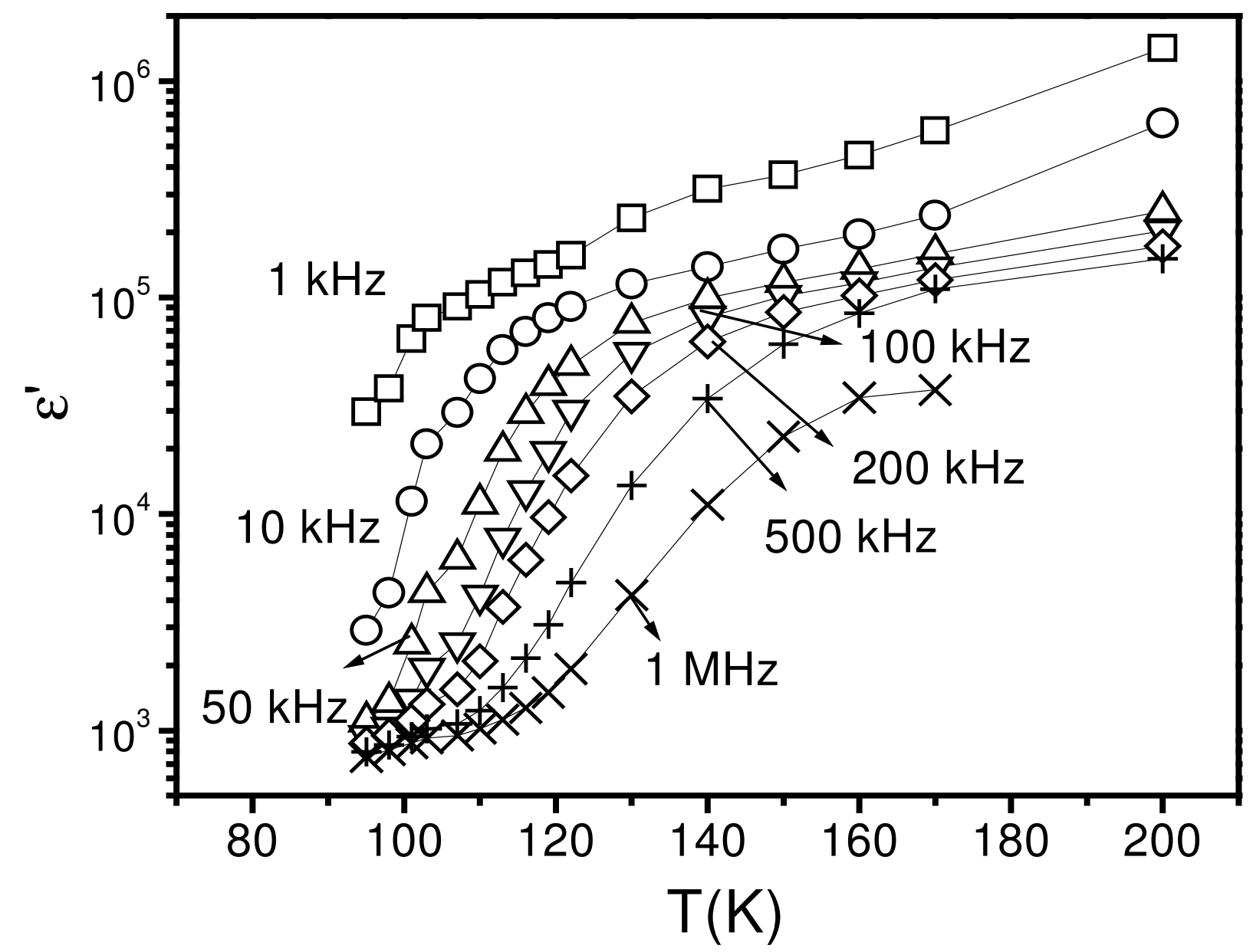

**Fig. 11** ***L. Sudheendra and C.N.R. Rao***